\documentclass[conference]{IEEEtran}
\usepackage{cite}
\usepackage{amsmath,amssymb,amsfonts}
\usepackage{algorithmic}
\usepackage{multirow}
\usepackage{graphicx}
\usepackage{textcomp}
\usepackage{xcolor}

\def\BibTeX{{\rm B\kern-.05em{\sc i\kern-.025em b}\kern-.08em
    T\kern-.1667em\lower.7ex\hbox{E}\kern-.125emX}}
\begin{document}

\title{Vulnerability Detection Through an Adversarial Fuzzing Algorithm\\
}

\author{\IEEEauthorblockN{Michael Wang}
\IEEEauthorblockA{
\textit{Montgomery Blair High School}\\
Silver Spring, MD \\
mxw206@gmail.com}

\and

\IEEEauthorblockN{Michael Robinson}
\IEEEauthorblockA{\textit{Department of Mathematics \& Statistics} \\
\textit{American University}\\
Washington, DC \\
michaelr@american.edu}
}

\maketitle

\begin{abstract}
Fuzzing is a popular vulnerability automated testing method utilized by professionals and broader community alike. However, despite its abilities, fuzzing is a time-consuming, computationally expensive process. This is problematic for the open source community and smaller developers, as most people will not have dedicated security professionals and/or knowledge to perform extensive testing on their own. The goal of this project is to increase the efficiency of existing fuzzers by allowing fuzzers to explore more paths and find more bugs in shorter amounts of time, while still remaining operable on a personal device. To accomplish this, adversarial methods are built on top of current evolutionary algorithms to generate test cases for further and more efficient fuzzing. The results of this show that adversarial attacks do in fact increase outpaces existing fuzzers significantly and, consequently, crashes found.

\end{abstract}


\section{Introduction}
Fuzzers are in widespread use for testing software. Used by hackers, cybersecurity professionals, software developers, and more, fuzzing is a valuable technique that discovers bugs and implementation flaws through immense swaths of malformed payloads. 

Fuzzing has been met with a lot of success in years past; the OSS-FUZZ \cite{ossfuzz} project revealed over 30,000 bugs in 500 open-source projects. Despite this, there remain many areas upon which existing fuzzers can be improved. In particular, the current state-of-the-art for software fuzzing has shifted towards an emphasis on maximizing code coverage. More recently, the application of machine learning has appeared in the field as well. Some notable examples of this include the use of RNNs in input generation \cite{godefroid}, the use of genetic algorithms to guide mutation strategies \cite{yigao} \cite{raydev} \cite{afl}, and the use of neural networks to model code coverage \cite{neuzz}.

In this article, an approach that combines fuzzing with adversarial attacks is explored. While traditionally a technique heavily employed in the machine learning fields of computer vision and image recognition, adversarial attacks find applications in many AI-gated systems \cite{cartella} \cite{irfanmuh} \cite{renzhang} \cite{chakraborty}. In this case, adversarial attacks could potentially position themselves well to assist fuzzing software due to its ability to carry out targeted attacks and transfer them to other systems \cite{tramrspace} \cite{szegedy}. As such, we develop the application of machine learning in fuzzing programs to improve the efficiency of existing fuzzing methods, even on low-performance devices. More specifically, we propose the augmentation of an adversarial method on top of a popular, existing evolutionary fuzzer, AFL (American Fuzzy Lop)\cite{afl}. We assert that our augmented method is extendable (with proper interfacing) to other software fuzzing programs and targets with slight modifications to existing code. In Section II, we discuss the implementation and methodology behind our approach. In Section III, we reveal the results of our method compared to the original fuzzer. Finally, in Section IV, we conclude with a summary of our findings and avenues for further research.

\section{Background}

\subsection{Neural Smoothing and Approximation}
The universal approximation theorem tells us neural networks are able to approximate any continuous function \cite{funahashi}. This is important as, while typical smoothing masks are built upon integrals and computed analytically \cite{neuzz}, it is infeasible to create a closed form of or analytically compute a computer program. As such, the use of a neural network could potentially be valuable in modeling program behavior by learning smooth approximations of their complex behaviors. This idea is demonstrated in several studies. While traditional software models have experienced success in their applications \cite{powell} \cite{connkatya}, they fall short on real-world programs with far more complicated behaviors. On the other hand, models that utilize similar machine learning methods prove to have far more success \cite{neuzz} \cite{saavedra} \cite{wangyanjia}. This is important in the context of our design because fuzzing requires the program to be able to explore, make sense of, and debug large, complicated programs.

Neural networks also are very relevant in the area of adversarial learning. The current most effective adversarial methods rely on the computation of gradients to search for adversarial samples. As such, adversarial methods have been almost exclusively reserved for deep learning models due to the ease of gradient computation and the lack of discontinuities in the generated approximation. A neural network therefore has the potential to prove itself as a strong choice in enabling the use of adversarial methods. It is not only capable of accurately capturing program behavior, but it attempts to create a smooth approximation of the behavior in the process. These properties enable the use of gradient-based attacks, which in turn could transfer to the program.

\subsection{Adversarial Attacks}

Adversarial methods have grown immensely due to the rising prevalence of deep learning. Adversaries can now manipulate and perturb DNNs to force instances of misclassification. But while this effect is well-studied in computer vision, this method draws a parallel to computer programs. By accurately representing a target program with a neural network, adversarial samples targeting the neural network can be crafted and by extension, samples that target the program itself. One particular method is chosen in the construction of the modified fuzzing algorithm within this paper, namely, the Jacobian Saliency Map Attack (JSMA) \cite{papernot}. 

\subsection{Fuzzing with AFL}
To understand the goal of the various methods presented in this fuzzer, it is important to understand how AFL maintains its process. Of the many components to AFL, it is particularly important to understand how inputs are mutated and then evaluated by the fuzzer.

For each input generated by AFL, there is a associated bitmap that stores the hits of each set of inputs encountered through the fuzzing process. This trace is compared to a global map updated throughout the entire fuzzing process to help AFL determine if any new behavior was found in the fuzzing process. Through this mechanism, AFL is able to track the branch (edge) coverage of the program, storing successful inputs as seeds to use as a starting point for future rounds of fuzzing. This will serve as the basis for our augmented method to find new areas in the program.

For the actual fuzzing, AFL's mutation algorithm can be split into two primary parts: deterministic and non-deterministic fuzzing. In the deterministic stage, the fuzzer iterates through the inputs and sequentially navigates through a series of preprogrammed mutations. These include bit flips, adding constant values, setting bytes to known-to-be-troublesome values, and more. As a consequence, whenever an input is fuzzed, AFL guarantees that the fuzzer searches within a certain area around the input. From there, AFL moves into its non-deterministic, also referred to as "havoc", stage. At this point, the fuzzer applies random tricks to mutate the input even further. These include operations such as input splicing, insertions, deletions, and more. As useful as this might be with some good luck, this also leads to many wasted mutations, inspiring the use of new methods to help the fuzzer land at more effective inputs more frequently.

\section{Methodology}

\begin{figure}[htbp]
\centering
\centerline{\includegraphics[scale=0.9]{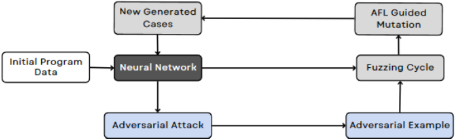}}
\caption{Summary of the pieces in the modified fuzzing algorithm.}
\label{fig4}
\end{figure}

The key idea behind the fuzzing scheme lies in the generation of adversarial samples to expedite the fuzzing process. Targeting inactivated edges, each test case is mutated following the computed gradient for the targeted edge. When combined with the blanket, deterministic fuzzing and random havoc stages performed by AFL, there is a higher chance of finding inputs that uncover rare edges.

\subsection{Preprocessing}

\begin{figure}[htbp]
\centerline{\includegraphics[scale=0.6]{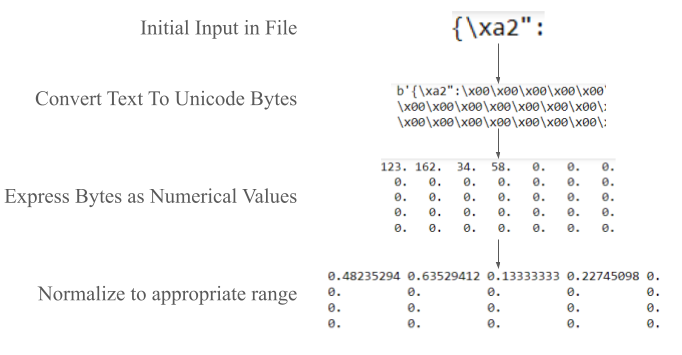}}
\caption{Walkthrough of the input encoding process.}
\label{fig2}
\end{figure}

The fuzzer starts from a set of program data, which involves an initial exploratory phase without a neural network in order to gather enough data. The data is then preprocessed, carefully tracking the number of times each edge appears across each test case. In order to encode the structured data in a way that is recognizable by both the fuzzer and the neural network, several additional steps are taken. Each input is converted to a series of Unicode bytes before being turned to a series of integer values and normalized. This process is summarized in Figure \ref{fig2}, where an example is given of the procedure.

Due to the limitations of a neural network, inputs into the network must always be set to a particular size. As such, the neural network is trained to accommodate the largest-sized input in the corpus, and all other entries are padded with null bytes. This helps to meet the size limitation without altering the input itself. 

The preprocessing of the output is relatively simple as well. In the execution of each test case, AFL returns a list of the edges encountered and assigns each one a unique ID label. With this, we compile each list together and return a one-hot-encoded vector, representing which edges a particular input encounters. There are also several additional steps taken during this stage to reduce the size of the output. Edges that are always activated together are condensed into one node to reduce the amount of information. Furthermore, only edges that have been activated at least once are included within the output vector to prevent overly sparse bitmaps.

\subsection{Neural Network}
In designing the neural network, a simple, feed-forwand dense neural network was used to increase the ease of gradient computation. As such, the network was kept relatively simple with only two hidden, fully connected layers of 256 neurons each. The network was also trained with binary cross-entropy as our primary mechanism to compute the distance between the predicted and true coverage of the outputted bitmap. Design-wise, because linear behavior in higher dimensions has been shown to increase the speed and effectiveness of adversarial attacks \cite{goodfellow}, the rectified linear activation unit (ReLU) is used in the hidden layers. On the outside, the network takes a fixed size program input in the form of a byte sequence. For its output, the network outputs an edge bitmap, essentially a one-hot encoded vector of all the unique paths discovered through AFL's fuzzing process. The output layer resolves to a sigmoid function because multiple edges can be traversed through one input. 

\begin{figure}[htbp]
\centerline{\includegraphics[scale=0.39]{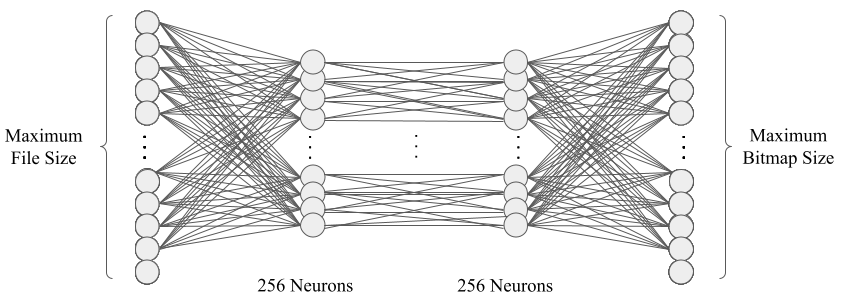}}
\caption{General architecture of the trained neural network. Consists of two hidden layers and trains to roughly 94\% accuracy each cycle.}
\label{fig1}
\end{figure}

The neural network is retrained and updated with new sample cases if any of the following three conditions are met. (1) A certain number of new cases have been found. (2) A new case is found that either discovers a new edge or is larger than any previous case. (3) The fuzzer has arrived at the end of a fuzzing cycle (All queue cases have been handled).

\subsection{Adversarial Method}
The method computes the signed gradient of the neural network with respect to the input in order to generate an adversarial saliency map. This method can find and leverage high-saliency features with an input, which are then mutated by the parameter that controls the magnitude of perturbation. These mutations are repeated up to a certain number of iterations with accordance to the JSMA attack. Note that although the JSMA attack is modified to meet the needs of program-based fuzzing, the essense of the algorithm remains the same. The attack constructs an adversarial saliency map - a input (in this case a vector) that ranks all the original input features chosen from how influential they are in leading the model to predict a certain class. In short, the algorithm consists of iterating the three following steps until either the target is reached or a certain number of iterations is reached. (1) Assuming \textbf{F}(\textbf{X}) to be the vector-valued function that maps to the output \textbf{Y} as computed by the neural network, compute $\nabla$\textbf{F}. (2) Construct an adversarial saliency map computed from the derivatives with respect to the input. (3) Modify the identified input features.

Under the native implementation, JSMA is configured to interfere with the final probabilities in a multi-class classification problem. However, computer software ends up as a completely different class of problem as each input has a set of edges it covers, making it difficult to set a constant adversarial target for each test case. While multi-label classification describes this situation, the fuzzer only cares about the probability of the target edge, meaning that the other labels only serve as extraneous information that may hinder the convergence of the attack. To address this, in this method, the probability returned through the network is preserved throughout the attack. Given that only one class is attacked at a time, this limitation can be worked around by ignoring the probabilities from non-target classes. Specifically, we first use a sigmoid function to evaluate the logits returned from the input. As in such a situation, each probability acts independently of the other, the success of the bitmap can be broken down and verified instead by separately checking each edge individually. As a consequence of this change, we preserve only the logits to the classes we are interested in targeting. In doing this, we remove any assigned importance on other edges, not only removing the need to read over extraneous edges, but also allowing the adversarial model to focus on our target classes, without the pressure to maintain or deactivate any activated paths from the initial input. This allows the attacks to converge more often onto target classes with higher confidence.

Another important alteration from the original JSMA is that the saliency computation doesn't rank inputs past the length of the input. While unlikely that these values are considered to be substantially salient by the algorithm, to be safe, we discount these values completely by removing them from the calculation. Instead, we substitute their values on the map with $0$, giving the minimum possible saliency value and removing any possibility of them being considered. The actual manipulation of the sizes of each input are handled by random insertions, deletions, splicing, and other strategies invoked through AFL. 

\subsection{Parallelization}

\begin{figure*}[htbp]
\centering
\hspace*{-3cm}  
\centerline{\includegraphics[scale=0.15]{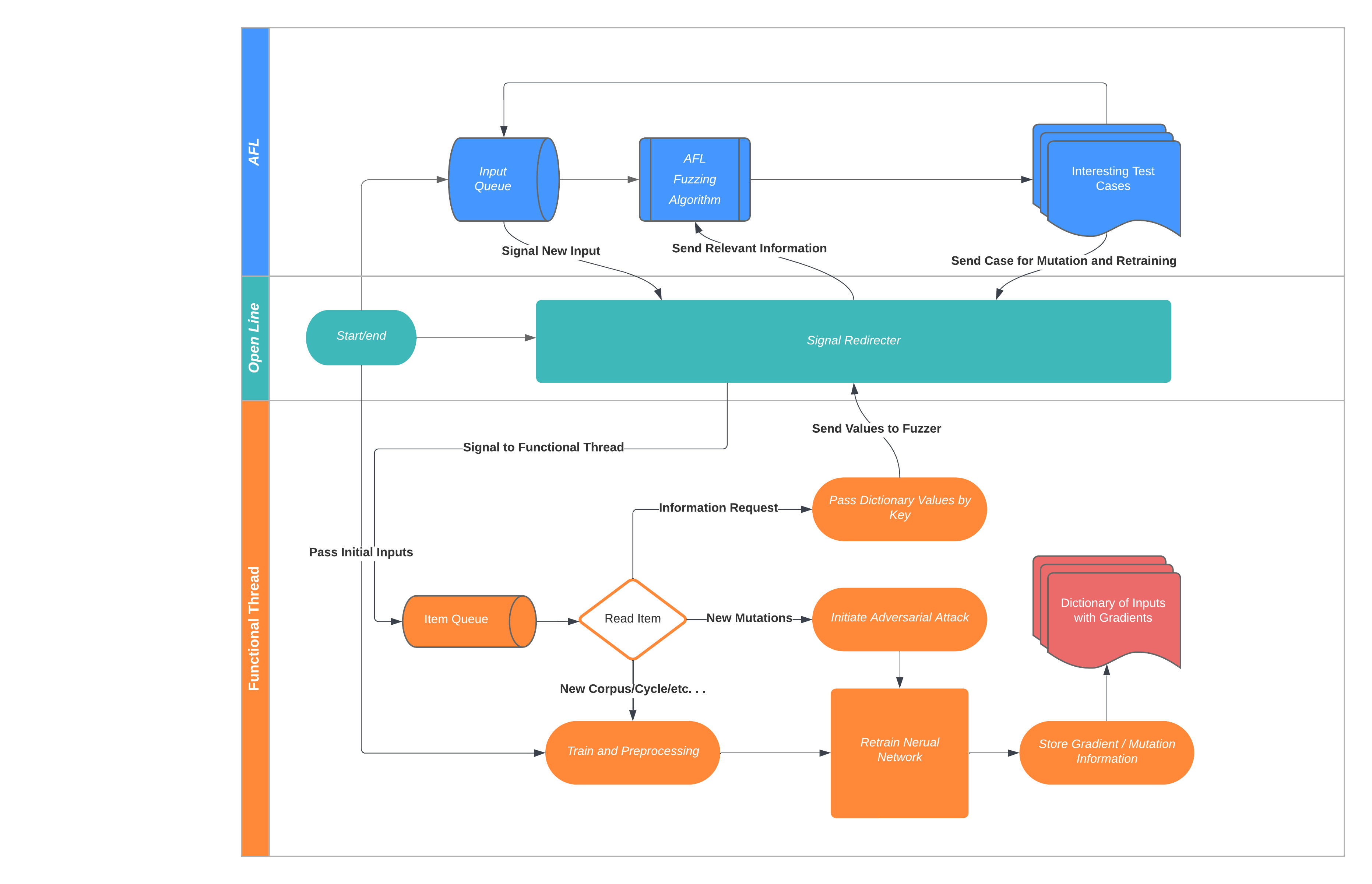}}
\caption{Diagram detailing how the parallel processes interact with one another.}
\label{fig10}
\end{figure*}

Both the neural network and adversarial method are run in parallel to the actual fuzzing process. As interesting test cases are passed into the fuzzing cycle, they are also passed into a separate process that crafts targets with adversarial samples targeting the $n$ least frequent edges depending on the program size. This operation is facilitated by splitting the fuzzer into two main processes. One process is the actual fuzzer, being AFL in this case. The other process manages the components of our augmented method, such as input processing, the neural network, and the use of adversarial attacks. Because our augmented method requires these the components to work closely together, we split the second process into two threads. One acts as a control tower for the functioning of the fuzzer, while the other remains open to log every signal it encounters. 

Combined, the two threads are set to listen and record all the communication done in a queue, which distributes commands to each of the two processes. Loading these commands into a queue is a relatively important part of this process because it ensures that instructions are sent in order, preventing the possibility of losing signals amidst the vast amount of information being communicated.

Between the fuzzer and these various components, sockets connections and file writing act as the primary forms of communication. Updated inputs discovered through the fuzzing processed are logged as text files in a folder which are then read by the preprocessing code. Likewise, mutated inputs are passed back into the fuzzing cycle for further mutation, either directly being sent via the socket, or stored as a file in the queue folder if deemed too large. Otherwise, various signals are sent via the socket throughout the fuzzing process. These often indicate key milestones such as new cycles, new paths discovered, and priority inputs to attack.

\section{Evaluation}
In this section, we discuss how the adversarial method was tested and how well the fuzzer performed.

\begin{table*}[ht]
\caption{Summary Statistics of the Four Tested Fuzzers On Fuzzgoat (5 x 30 Minute Trials)}
\begin{center}
\begin{tabular}{cccccc}
\hline
\textbf{Evaluated}&\multicolumn{5}{c}{\textbf{Metrics}} \\
\textbf{Fuzzers} & \textbf{\textit{Average Executions Per Second}}& \textbf{\textit{Total Executions (Millions)}}& \textbf{\textit{Levels of Mutation}}& \textbf{\textit{Paths Found}}& \textbf{\textit{Unique Crashes Found}} \\
\hline
Adversarial & 5078.26 & 7.73 & \textbf{21} & \textbf{699} & \textbf{46}\\
AFL & 5522.47 & 8.48 & 26 & 618 & 31\\
AFL++ & \textbf{6004.86} & \textbf{9.07} & 31 & 646 & 40\\
FairFuzz & 4749.93 & 7.51 & 25 & 673 & 38\\
\hline
\end{tabular}
\label{tab1}
\end{center}
\end{table*}

\begin{table*}[ht]
\caption{Edge Coverage of the Four Tested Fuzzers On Other Programs (3 x 24 Hour Trials)}
\begin{center}
\begin{tabular}{c@{\hskip 0.75in}c@{\hskip 1in}c@{\hskip 1in}c@{\hskip 1in}c@{\hskip 0.35in}}
\hline
\textbf{Evaluated}&\multicolumn{4}{c}{\textbf{Targets}}{\hskip 0.6in} \\
\textbf{Fuzzers} & \textbf{\textit{libxml2-v2.9.2}}& \textbf{\textit{libpng-1.2.56}}& \textbf{\textit{openssl\_x509}}& \textbf{\textit{mupdf}} \\
\hline
Adversarial & 5154 & 1713 & 1255 & \textbf{417} \\
AFL & 5002 & 1690 & 1144 & 360 \\
AFL++ & \textbf{5441} & 1727 & \textbf{1298} & 388 \\
FairFuzz & 4688 & \textbf{1750} & 1287 & 355 \\
\hline
\end{tabular}
\label{tab2}
\end{center}
\end{table*}

\subsection{Target}\label{AA}
This fuzzer was tested across a variety of open source programs of different sizes and input formats in order to thoroughly evaluate the performance of the fuzzers. The programs are Fuzzgoat\cite{fuzzgoat}, libxml2, libpng, openssl\_x509 and mupdf. While the other four are popular real world programs, thus making them valauble to test, Fuzzgoat was chosen for a different reason.
Fuzzgoat is a deliberately flawed C program of decent length made to test fuzzers. This choice was made to test the mechanism in an environment where vulnerabilities were already marked in the code for analysis. Additionally, Fuzzgoat could execute relatively quickly, meaning that differences in computational efficiency are more transparent, as differences in runtime would not be related to the load incurred from the program.

A total of four fuzzers, the modified algorithm, AFL \cite{afl}, FairFuzz \cite{fairfuzz}, and AFL++ \cite{AFLplusplus}, are tested and compared with one another to evaluate how this fuzzing method shapes up to the industry standard. AFL serves as a base comparison for the fuzzer as the augmented method is built on top of the AFL algorithm. FairFuzz was also identified as a valuable fuzzer for comparison because like the adversarial method, it attempts to target rarer branches while adopting a far different approach. Finally, AFL++ was included as a way to compare this method to what is widely regarded as the industry standard. Each test was an average of five runs on Fuzzgoat for 30 minutes each, which proves to be enough time to explore the majority of the program, regardless of which method we employ. For the other programs, the results are taken from 3 runs of 24 hours each. 

The fuzzers are run on a machine with an AMD Ryzen 7 5700U processor, 12 GB memory, 64-bit Microsoft Windows 11 Home Operating System, and x64-based processor. Each process is hosted on Ubuntu 20.04.4 LTS with 5888 MB of memory each.

\subsection{Results}

We present three primary differences in the performance of the modified fuzzing algorithm:

\subsubsection{Runtime}



For this section we look exclusively at Fuzzgoat as is a much more controlled environment. Any differences in runtime/computational efficiency are more noticeable because we reduce the effect of any load brought by the program itself. Notice in Figure \ref{tab1} that the execution speed of the original AFL is notably higher than the execution speed of the adversarial fuzzing algorithm. Specifically, the execution speed experienced by AFL is on average, $5522$ executions per second, whereas the execution speed of the adversarial fuzzing algorithm falls to about $5078$ executions per second; A near $10\%$ difference in execution speed. This is explainable as the revised fuzzing algorithm is more computationally intensive and contains many more parts. Hardware aside, the adversarial algorithm relies on thousands of rapid gradient computations to arrive at a new test case, something that AFL does not have to do. The added workload likely inhibited the fuzzer, especially since the processes were run in the same instance on the same machine. Despite this, the adversarial fuzzing algorithm can keep up with the other two fuzzers in terms of speed, even beating FairFuzz. This is reflected in Table \ref{tab1}, where AFL++ ran the most executions, followed by AFL, then the adversarial method, and finally FairFuzz, at $9.07$, $8.48$, $7.73$, and $7.51$ million executions respectively. Throughout the fuzzing process, the adversarial fuzzing is able to consistently maintain its performance, indicating its performance is comparable to simpler genetic fuzzers that lack complicated machine learning architecture.

\subsubsection{Path Coverage}

\begin{figure}[htbp]
\centering
\centerline{\includegraphics[scale=0.35]{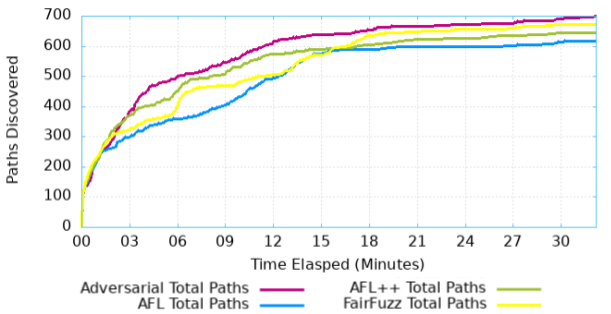}}
\caption{Comparison over time between the path coverage of the adversarial fuzzer, AFL, AFL++, and FairFuzz when tested on Fuzzgoat.}
\label{fig6}
\end{figure}

The results in Figure \ref{fig6} indicate that the adversarial fuzzer accomplishes its task when compared to AFL. Looking only at Fuzzgoat, despite running slower than the other algorithms, the coverage exhibited by the algorithm still finds far more paths in the same amount of time. Furthermore, the adversarial fuzzer also leads the other fuzzers while fuzzing Fuzzgoat, showing how it can find these paths sooner than other methods. Notably, FairFuzz, AFL++, and AFL dig significantly deeper into their test cases to find the same results. On average, in 30 minutes FairFuzz, AFL++, and AFL explore $25$, $31$, and $26$ levels, respectively, while the adversarial algorithm searches only $21$. This difference suggests that adversarial methods could successfully inject much more valuable test cases to fuzz, skipping over extra levels of mutation spent searching for critical inputs. 

\begin{figure}[htbp]
\centering
\centerline{\includegraphics[scale=0.5]{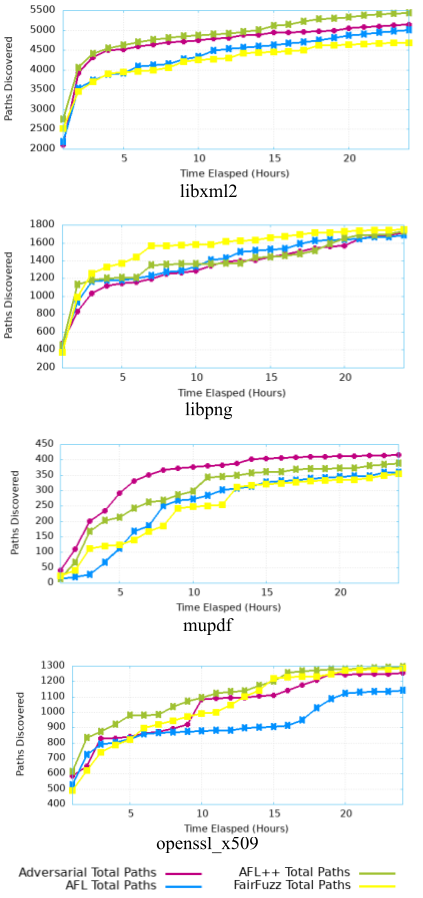}}
\caption{Comparison over time between the edge coverage of the adversarial fuzzer, AFL, AFL++, and FairFuzz when tested on various real world programs.}
\label{fig11}
\end{figure}

That said, when we compare the performance of of each fuzzer across our various real world targets, these results become slightly different (see Figure \ref{fig11}). Over longer periods of time, the edge coverage with respect to time tends to even out,  While AFL++ boasts the highest edge coverage overall, the adversarial method seems to follow behind, outforming both AFL and FairFuzz. This indicates the adversarial methods are successful in deriving cases that produce extra edge coverage when compared to its original algorihtm (AFL). Notably, the performance of our adversarial method seems to excel in medium to smaller sized programs like mupdf and Fuzzgoat, while AFL++ is able to experience more success with larger programs like libxml2 and opennssl. When compared directly to AFL, we can see that the adversarial fuzzer is able to consistently find more paths and in shorter aounts of time as well. For instance, while AFL struggled with openssl, the adversarial algorithm was able to overcome this issue and remain on par with the other two fuzzers. Furthermore, in mupdf and libxml2, the adversarial algorithm showed the ability to discover new areas sooner than AFL.

\subsubsection{Crashes}



Crashes found are an interesting metric because detecting bugs in the ultimate goal in softare fuzzing. That said, with the exception of Fuzzgoat, crashes found is a poor metric to evaluate with because bugs are typically very sparse in real world programs. As such, this section will primarily focus on the fuzzing results from Fuzzgoat in order to provide evidence that edge coverage is a valid proxy for bugs discovered. It can be observed in Figure \ref{tab1} that crashes are found much sooner and faster in the adversarial scheme than in the other three fuzzers. The adversarial algorithm uncovers $46$ instances of unique crashes while FairFuzz, AFL++, and AFL find $37$, $40$, and $31$, respectively. This aligns with the findings for path coverage. It is reasonable to link coverage and crashes together as with more paths discovered, the bugs within those paths are discovered as well. While not trained to look for harmful inputs, the results show that by increasing path coverage, vulnerability detection also increases. We suspect that the adversarial method can do this because the adversarial method activates a far greater area in the input space. This is in accordance with previous studies that find in their applications that path coverage increases bug detection \cite{neuzz} \cite{yigaolei} \cite{zhaoxiaoqi}.

\begin{figure*}[htbp]
\centering
\centerline{\includegraphics[scale=0.45]{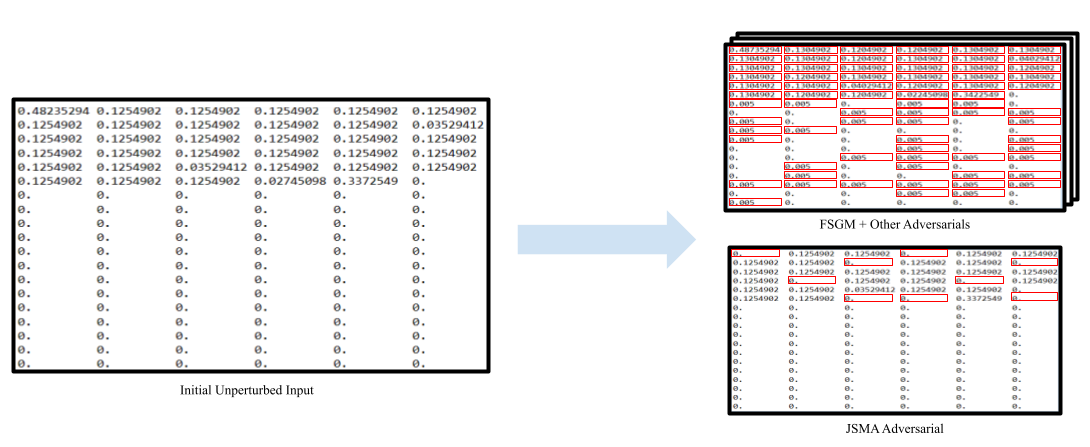}}
\caption{Examples of a JSMA and FSGM attack run on an example input discovered during the initial discovery phase of the fuzzer. Notably, the JSMA only mutates a few select features (boxed in red), while FSGM affects almost the entire feature space to achieve the same result.}
\label{fig3}
\end{figure*}

\section{Discussion}

The results of our method show that this technique is a promising way to increase fuzzing efficiency. With a neural network, we can effectively identify points of attack and engineer specific edges to appear by conducting a saliency map attack. However, along with this, there are a few things to note.

\subsubsection{Model Performance}
Model performance has proved to greatly affect the fuzzer. As the algorithm is based on a gradient-guided optimization method, the return gradients may not be accurate, which is a problem only exacerbated when the neural network itself is not accurate. Thus, we see a fall in performance when we train the network inadequately.

\subsubsection{Hardware}
While the results of this study show promising results in overcoming challenges in computation efficiency, the more powerful, the better. This experiment was not run on an optimal configuration, and because of that, the model may have suffered in several ways. Notably, the speed at which adversarial inputs can be generated and inserted into the fuzzing queue slows significantly.

This has several important ramifications. The current attack is configured to search for the first 50 uncommon edges, primarily because each attack takes time, and many seeds need to be mutated. In addition, the amount of iterations the attack takes has been reduced from 2000 to 200. This change is also to speed up each case and is just another instance where we sacrifice some adversarial samples so that the attacks can keep up with the fuzzer. Given this, it is reasonable to believe that the results of this fuzzing scheme would only be better with different equipment. This difference would allow for an increase in not just the quantity of samples generated, but the quality of each one as well. Attacks could search more rigorously and identify potential samples that exist further away from the seeded input. Similarly, the reduced execution speed on our adversarial method could be mitigated if run on separate instances. Despite this, this fuzzing scheme still shows impressive adaptability, especially when compared with other state-of-the-art fuzzers in Tables \ref{tab1} and \ref{tab2}.

\subsection{Adversarial Method}

When comparing this attack to others, we find that this mechanism fits a few important criteria which makes it superior to other forms of gradient-based attacks.

\subsubsection{Null Byte Spillover}
One of the unique features of JSMA is its adversarial saliency map. The map identifies a set of inputs most important in mutating the output in its desired direction. Using this mechanism, we can specifically craft saliency maps that focus only on features that already exist within the current input. This identification helps prevent any spillover into null bytes. We can observe this difference when comparing JSMA with other methods. Although methods like the Fast Gradient Sign Method (FGSM) \cite{goodfellow} and Projected Gradient Descent \cite{kurakin} create small perturbations, the noise gets distributed across the entire input, often activating many null bytes by accident. This is demonstrated in Figure \ref{fig3}. As such, inputs are generated using a saliency map to avoid this issue. 

\subsubsection{Perturbation Magnitude}
As research into adversarial attacks continues, there is a continued emphasis on the imperceptibility of attacks. While useful in the context of image classification and the like, that is less of a concern in fuzzing. Popular attacks like the Carlini and Wagner attack \cite{carlini2017}, optimize to minimize distance metrics, which is not particularly important in this context. On the flip side, JSMA can be configured to make larger, more meaningful mutations, which usually get split into numerous, smaller perturbations in other attacks (see Figure \ref{fig3}). Furthermore, saliency map attacks have a great degree of freedom for distortion and feature variation, which allows the algorithm to set bounds on how much an input changes and how much to modify each input feature.

\subsubsection{Guided Mutation}
The use of a saliency map is particularly beneficial in mutation-based fuzzing algorithms. In identifying groups of important features, the algorithm can focus on mutating only high-impact bytes, which is apparent in Figure \ref{fig3}. Furthermore, because the algorithm iterates over these gradients many times, we can arrive at inputs that would take multiple rounds of fuzzing to reach. These two factors greatly decrease the amount of potentially wasteful mutations.

\section{Conclusion}
This report presents a new fuzzing scheme that combines the concept of adversarial attacks with the fuzzing of a target program. Unlike current GAN-related approaches with fixed-sized outputs generated between fuzzing cycles, this fuzzer implements a version of the Jacobian Saliency Map Attack to craft samples to preserve the length of original inputs and run parallel to the fuzzing process. Generated cases are inserted into the fuzzing queue to be fuzzed live and new interesting cases are sent back to the network to refine the model. An evaluation of the setup indicates that this method increases the coverage of AFL and by extension, the bugs found within target programs. This method was also compared to other common fuzzers on real world programs and showed comparable performance. The findings further suggest that this method could be even more powerful if done with improved hardware setups, more efficient fuzzing/coding practices, and optimized parameter configurations.

As research into adversarial techniques and neural networks progress, we can expect to see a more mathematically concrete version to optimize the susceptibility of the network while minimizing the impact on accuracy. The closer we get to a solution, the easier it is to find adversarial examples in space, greatly enhancing the potential of this methodology.

\section*{Acknowledgment}

We acknowledge the The Blair Jones Fund from the American University Mathematics and Statistics Department for hosting this project and making it possible.

\bibliographystyle{IEEEtran}
\bibliography{bibliography}

\vspace{12pt}

\end{document}